# The teacher of the *gattini* (kittens)

R. Casalbuoni, D. Dominici

Abstract: *The figure of Raoul Gatto, who died in 2017, is remembered here, with an illustration of his career and the main results of his research, along with the personal memories of some of those who worked with him.*

Raoul Gatto, a great master of Italian theoretical physics who had scientifically trained numerous students, the so-called *gattini,* meaning kittens (gatto being the Italian word for cat), and made significant scientific contributions to the progress of theoretical physics, died in Geneva on the 30$^{th}$ of September 2017.

Raoul Gatto was born in Catania on the 8$^{th}$ of December 1930. His father was an engineer and his mother a housewife. After his secondary school studies at the *Liceo Scientifico G. Oberdan* in Trieste, where other well-known physicists studied too, he seemed destined to study engineering but, in 1946, he won the competition to enter the *Scuola Normale Superiore* in Pisa and, in 1947, he began a bachelor's degree course in Physics. Gatto said that his years at the *Normale* provided great intellectual stimulus, especially in relations with students of the Humanity Faculty, some of whom became famous writers such as Pietro Citati and Carlo Sgorlon. In 1951, he graduated with a first class honours degree (110 cum laude), with a thesis on shell nuclei models, under the guidance of Bruno Ferretti, who occupied the chair of theoretical physics in Rome at the time, and Marcello Conversi, who, teaching experimental physics in Pisa, was his mentor. After his thesis, Gatto moved to Rome, where he became Bruno Ferretti's assistant, devoting himself to the study of the weak decay of hadrons and their associated angular distributions. In Rome, Gatto found a very open international environment and remarkable scientific stimuli, thanks to Edoardo Amaldi and the legacy of the Enrico Fermi school.

In particular, Gatto suggested the preservation of CP symmetry (a product of the conjugation of charge and parity), independently of Landau and Lee and Yang. Fitch and Cronin's 1964 experiment showed that CP symmetry is also violated.

Upon returning to Italy in 1960, Gatto became professor at the University of Cagliari where he remained until 1962, commuting with the Frascati Laboratories of the National Institute of Nuclear Physics. In the spring of 1959, at the Frascati Laboratories and under the direction of Giorgio Salvini, the 1.1 GeV electro-synchrotron came into operation, bringing Italian Physics to the frontier of energy at international level, there being only two other accelerators in the world at that time comparable in type and size, one at Caltech and one at Cornell. In March 1960, Bruno Touschek held a seminar in Frascati on the importance of a systematic study of the collisions between electrons (e-) and positrons (e+) and on how these could be obtained with electron and positron beams circulating in opposite directions in the same ring, using just one magnet. This was the launch of the AdA (*Anello di Accumulazione* or Accumulation Ring) project, the first e+e- collision device in the world with a 220 MeV energy beam, and therefore 440 MeV in the mass centre system. AdA was to be followed by the ADONE project, another e+e- collider, with about 3 GeV energy in the centre of mass.

In 1960 Gatto began a collaboration with Nicola Cabibbo, who graduated in 1958 with Touschek, to study the phenomenology of high-energy positron electron collisions. The physicists of Frascati called this article The Bible, due to the completeness of its content for potential applications to AdA and ADONE. With Cabibbo, Gatto also wrote a work in which SU(3) symmetry was used in weak interactions. This work formed the basis of the work in which Cabibbo introduced the angle that later became known as the Cabibbo angle. From the time of its creation in 1960 until 1964, Gatto was also Head of the Theoretical Group of National Laboratories of Frascati.

An amusing episode from that time is remembered by Mario Greco and Giulia Pancheri [Greco, 2008]: "At the time, in order to write a thesis on theoretical physics, you had to sit a short test, solving a problem assigned by one of the members of the theoretical group. The best students of the time were involved. And so it was that Gatto's two graduands, Guido Altarelli and Franco Buccella, appeared in Frascati. The latter was known as "the other graduand" because of a prank played behind their backs by Gianni De Franceschi, a young theorist at Frascati who dealt with group theory. De Franceschi had graduated with Marcello Cini and arrived in Frascati with a scholarship in September 1960. He had then started collaborating with Gatto and Cabibbo. De Franceschi was very good at imitating voices and one day, in Frascati, he called Altarelli from the telephone in the next room and, pretending to be Gatto, gave him a thorough telling off, accusing him and the "other graduand" of not working enough and of being slackers.".

In academic year 1962/63, Gatto moved to the University of Florence (see Fig. 1) to cover the chair left free by Morpurgo. The scientific context in which Gatto found himself is illustrated here, with Table 1 indicating the composition of the research groups of Florentine physics at that time.

Table 1
**Research groups present in Florence in 1962 in the four Institutes of Physics, of Theoretical Physics, of Astronomy and of *Fisica Superiore***
Theoretical physics: problems with many bodies, particularly the study of liquid helium. Simone Franchetti, Antonio Mazza
Experimental physics: low energy nuclear physics. Manlio Mandò, Tito Fazzini, Mario Bocciolini, Pier Giorgio Bizzeti, Anna Maria Bizzeti Sona, Pietro Sona, Giuliano Di Caporiacco, Marco Giovannozzi
Experimental physics: research into the physics of high energies with photographic emulsions. Michele Della Corte, Maria Grazia Dagliana, Anna Cartacci, Giuliano Di Caporiacco

Theoretical physics of elementary particles. Raoul Gatto, Marco Ademollo, Claudio Chiuderi, Giorgio Longhi
Didactics: Marco Giovannozzi
Microwave Centre: Nello Carrara, Giuliano Toraldo di Francia
Astronomy: Giorgio Abetti, Guglielmo Righini, Giovanni Godoli, Maria Cristina Ballario, Mario Rigutti, Giancarlo Noci

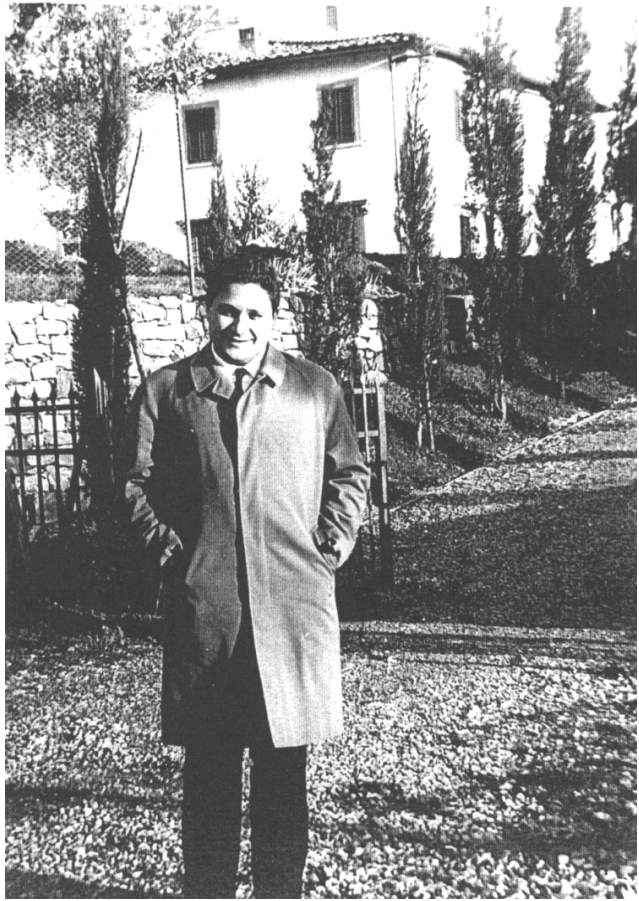

Fig. 1 - *Gatto during the Florentine period*

A year after his arrival, on the 19$^{th}$ of November 1963, Gatto was appointed president of the *Scuola di Perfezionamento* and in December of the same year, he managed to obtain five scholarships with a value of one million Italian Lira, to allow a year of attendance of the school by Emilio Borchi, who had graduated in Florence, and by Enrico Giusti, Guido Altarelli, Franco Buccella and Giovanni Gallavotti, who had graduated in Rome with theses on Electrodynamics in processes relevant to accumulation rings. This is how Gatto himself remembered that time: "In a certain sense, Florence was my first university. In Cagliari, although I had tried, it had been impossible for me to create a consistent group (isolation, total absence of INFN, financial poverty). Many young people in Rome had had scientific relations with me, and when they learned of my move to Florence they made it known that they wanted to follow me. But there was virtually no funding available. Luckily, (Marco)

Ademollo, (Giorgio) Longhi, (Claudio) Chiuderi, (Emilio) Borchi, (Mario) Poli and (Giovanni) Martucci were there, and I was happy to work with them, despite the fact that I didn't know them. Almost all newcomers came with financial means from their offices or with other types of funding... This made it impossible to keep them in Florence for long. The group's financial base was therefore weak and transitional. Luckily, Cabibbo had an American contract that allowed him, in the end, to take some at least for a short time. A typical example is that of (Luciano) Maiani. Luciano had completed a thesis on experimental physics with (Mario) Ageno at the Ministry of Health. But he decided to move on to theoretical physics. Ageno was a friend of mine. He called me in Florence and asked me if I could take Luciano, whom I had never seen. I trusted Ageno's scientific intuition, but told him that I had no money. And Maiani was sent to Florence. Of course, Arcetri suffered tremendous setbacks. ... Arcetri remained in total darkness in the evening and at night. But the boys (and I myself) also wanted to work at night. The battle for night lights was tough, due to opposition, for example, by Sassolini (the caretaker, ed) who was worried about security. Franchetti helped me and had the brilliant idea of telling Sassolini and the others who shared his negative opinion that lighting would actually discourage those with ill intentions. So we could work at night too. (Answers by R. Gatto to some of Casalbuoni's questions).

And again: "Marco Ademollo was one of the most continuous collaborators within the group. It's interesting that one of Marco's initial interests was related to a problem with formal aspects that were somewhat taxing but which, in a sense, remained marginal: the problem of spin determination. In fact, at that time, discoveries of new particles and resonances followed one another ceaselessly, and rapid and complete methods were needed to determine the spin, starting from various angular correlations. Giuliano Preparata later joined this line of research. Marco's interest later shifted, as it did for the whole group, to the properties of symmetry. The non-normalisation theorem was obtained using an algebraic method (known as the Ademollo-Gatto Theorem, ed). Collinear subgroups were one of the subsequent interests. Franco Buccella was initially interested in this line. Later, in addition to Ademollo, Longhi and (Gabriele) Veneziano showed an interest too. Also of interest at that time were sidewise dispersion relations, on which Ademollo's work was followed by the activity of (Roberto) Casalbuoni and Longhi. Longhi's activity was long and constant. In addition to what has already been mentioned, he dealt at length with the electronic positron radiation corrections. Altarelli and Buccella had already dealt with the subject in their thesis in Rome. Longhi worked on it with Altarelli, (Enrico) Celeghini, and (Silvio) De Gennaro. Pino Furlan also collaborated in this research.

Also on electron-positron, Longhi dealt with barionic resonances, together with Casalbuoni. The topic had already been dealt with by Celeghini, after his work on $\eta_0$. Celeghini was one of the first to come to Florence from outside, and later participated in the work on the saturation of the algebra of the currents.

A special mention should go to Borchi's activity, especially for his work on what was the first classification of mesonic resonances obtained by guessing that these resonances were part of a p-wave multiplet. The foundation of this classification, which was then very speculative, has proved to be correct over time and continues to be accepted, as stated in recent works by 't Hooft, Maiani, et al. Maiani, Preparata, Longhi and Veneziano furthered these studies shortly after. At that time there was considerable uncertainty regarding mesonic resonances. It was a question of sorting through Particle Data and seeing if you could make an attempt at classification. The introduction of p waves seemed necessary. It should be mentioned that Borchi had also contributed to work on muons (radiative capture) by studying the possible use of data to verify the properties of weak currents. A notable activity of the group was dedicated to the search for properties of symmetry of strong interactions, trying to exploit the phenomenological success of SU(6), which then seemed difficult to formulate on a theoretical level. Altarelli, Buccella, Maiani and Preparata were at the forefront in participation in this theoretically complex effort.

Buccella and Veneziano became interested (together with Okubo) in the thorny problem of Schwinger's (then almost unknown) terms.

Also worthy of mention is Altarelli and Buccella's interest in the non-leptonic decay of hyperons, which led at that time to suggest relations between amplitudes, later proved correct, despite the uncertain theoretical foundation SU(6).

Gallavotti, who possessed a vast mathematical culture, wanted to tackle the problems of statistical mechanics and clarify delicate matters, such as the structure of the invariance of gauges in Electrodynamics. Along with him, a mention should go to the presence of people with a clear mathematical tendency, such as (Enrico) Giusti, (Umberto) Mosco and, above all, (Giuseppe) Da Prato. The mathematical interest of these young men came from theoretical physics and their participation in discussions and seminars was constant. They went on to have brilliant careers in mathematics." (Gatto speaking about the theoretical physicists of the Florentine period). At the Europhysics conference of 1968, Harry Lipkin coined for this Florentine school created by Gatto the name of new Florentine Renaissance.

Gatto was not seen much around the Institute of Physics. He mostly stayed shut away in his room to work. Next to his door, in the hallway, there was a lamp with two colours of light, red and green. You could only enter his office if the light was green. He regularly received his assistants to take stock of the work they were doing. Because of these meetings, which sometimes lasted a long time, Gatto was often very late for lessons, a habit not appreciated by his students who had to endure lessons continuing until well after 2.00 p.m. and were forced to make considerable efforts to follow the explanations despite being starving.

In 1966, Gatto decided to enlarge the group and began a recruitment campaign among the third-year students, sitting in on the exams of Institutions of Theoretical Physics, a course held by Ademollo, while Gatto taught Theoretical Physics, and offering the students who seemed most promising the chance to do a thesis in Theoretical Physics. This is how Roberto Casalbuoni, Luca Lusanna, Paolo Gensini, Antonio Conti and Renzo Collina were recruited. Emanuele Sorace was already working on a thesis with Franco Buccella. It was very important to Gatto that his students be in the best position to study and prepare their thesis. He decided to use a corridor on the first floor of the Institute to have Rossi (the Institute's carpenter) build six wooden cubicles. These cubicles were then furnished with a desk, a small bookcase, a table lamp and an electric heater. These tiny "offices" were used in the years that followed by all the theoretical students.

After taking leave from 1-10-1967 to 30-9-1968, Gatto moved to the University of Padua in the 1968/69 academic year. In Padua he began working with Giuseppe Sartori and

Guido Tonin, writing a work in which the possibility of a connection between the Cabibbo angle and quark masses was suggested for the first time.

The thesis examination for Casalbuoni, Lusanna, Gensini, Conti and Collina took place on the 20th of March 1969 and Gatto attended the examination as supervisor for the theses of Casalbuoni and Lusanna. Lusanna had to leave shortly after for military service, but Gatto immediately set to work for Casalbuoni, assigning him an INFN position of temporary R6, starting from the first of April of the same year, pending the announcement of ministerial grants for the following year.

Longhi had also collaborated as a co-supervisor for Casalbuoni's thesis on the saturation of the algebra of currents using relativistic equations with an infinite number of components. This thesis generated a collaboration with Gatto that continued until 1971, also favoured by the fact that Casalbuoni had obtained a scholarship which he spent in Padua in 1970.

In 1971, Gatto moved to the University of Rome where, together with Sergio Ferrara, Aurelio Grillo and Giorgio Parisi, he wrote important works on the conformal group and, in particular, the two-dimensional conformal anomaly. In 1975 he began a collaboration with Riccardo Barbieri, Reinhart Kogerler and Zoltan Kunszt on the calculation of the widths of mesonic resonances with a confining potential, a project that continued after Gatto moved to the University of Geneva in 1976/77 to occupy the prestigious chair previously held by Stuckelberg.

Once in Geneva, Gatto did not form a theoretical group as he had done in the Florentine period. Instead, he used to invite many visiting researchers to the Department of Theoretical Physics in Geneva for varying periods of time. In the beginning, he continued some of his previous collaborations with Barbieri, Kogerler, Kunstz, Sartori, Vendramin and Celeghini, but then also began new ones with Franco Strocchi and Giovanni Morchio, from Pisa, and Francesco Paccanoni, from Padua. In Geneva Gatto had an assistant, Carlos Savoy, a Brazilian researcher who was in Padua when Gatto was Professor at the university. Savoy and Mario Abud (a PhD student in Geneva) collaborated with him at that time.

In 1979 Gatto resumed his collaboration with Florence thanks to a three-month stay by Casalbuoni at CERN. After a seminar by Haim Harari on the rishon model, a model in which quarks and leptons appeared as states composed of new particles, rishons, Gatto and Casalbuoni, while discussing the results, came up with the idea of using fermionic oscillators for a unified description of quarks and leptons. This was the new beginning of a collaboration that was to last until 2006.

In 1980 Gatto began a collaboration with Michele Caffo and Ettore Remiddi, both from Bologna, and with Barbieri on strong radiative corrections to the decay of quarkonium.

Gatto's collaboration with Casalbuoni on composite models for quarks and leptons was extended in 1982 to Francesco Bordi, who graduated with Casalbuoni, and Daniele Dominici, who had done his thesis with Longhi. These studies ended in 1983 with the proposal of a supersymmetric composite model. This line of research aimed at determining explicit models ended when interest shifted to the indirect effects of this kind of physics, possibly independently of the models themselves.

In the wake of the thesis on the dynamical breaking of symmetry in QCD discussed, in 1983, by Stefania De Curtis under the guidance of Casalbuoni, many works were begun in collaboration with Gatto, ending in 1988 with a work on the determination of the mass of quarks that was cited by the Particle Data Group as one of the estimates, existing at that time, of these important parameters. Andrea Barducci, De Curtis, Casalbuoni and Dominici participated in this research. Although the style with which Gatto led this group was different from that used in Florence, there began to be a substantial group of Florentine researchers who gravitated to the Department of Theoretical Physics in Geneva, staying for varying period of time.

Stefania De Curtis recalls "My first period at the University of Geneva was immediately after the discussion of my thesis. We had derived a modified version of the action for composite operators to study the dynamical symmetry breaking in theories with confining strong interaction, and Raoul was very interested in the subject. He asked me to explain some details of the derivation. I was fresh from "calculations", so I wasn't intimidated. We spent a whole afternoon at the blackboard, Raoul wanted to know everything, asking specific questions, and I, knowing the answers, was, let's say, proud to show him the results we had obtained with Roberto. I later learned that Raoul was impressed by my attitude and had asked Roberto if I was a member of the feminist movement. I never understood the link between the dynamical symmetry breaking and feminists. Probably Raoul was not used to working with women!

Whatever, this earned me a place among the *kittens*. Raoul later proposed that I do a doctorate at the Institute of Theoretical Physics at the University of Geneva. I opted instead for SISSA, but our collaboration continued, on various topics, for more than 20 years."

A break in the QCD line of research occurred in 1985. The Standard Model was now fully accepted, especially after the discovery of the vector bosons W and Z. Apart from precision checks, for which the beginning of the LEP (Large Electron Positron collider) operation at CERN was awaited, the missing element (in addition to the quark top) was the Higgs boson, on which the mechanism of electro-weak symmetry breaking rested. Given the lack of experimental information on the matter, the idea began to circulate that the breaking mechanism was more complex, with the possibility that it was due to a strong interaction on the TeV scale, similar to QCD, called technicolor. Again, interest revolved around the possible low-energy effects of such an interaction. Gatto and the Florentine group, Casalbuoni, De Curtis and Dominici, formulated an effective theory for such a scenario, which envisaged the existence of new vector bosons. This collaboration was facilitated by Dominici's stay at the University of Geneva for two years. Meanwhile, Gatto continued to collaborate with Sartori on the symmetry breaking, and with Caffo and Remiddi on the physics that could be explored at LEP.

Studies on Higgs continued for a long time. In addition to the phenomenological study of the model that had been introduced by the collaboration, the so-called unitarity limits that allowed the derivation of theoretical limits on the mass of the Higgs were also investigated. Carlo Giunti, another student of Casalbuoni, also participated in this research, as did Ferruccio Feruglio from 1987, having arrived in Geneva that year as a PhD student of Gatto.

In 1987, Casalbuoni won a place as professor and was called to Lecce. Since teaching involved him only for a semester, until he was called back to Florence in 1990, he used to spend the rest of his time as a visitor at the University of Geneva. At that time the group was working at full capacity, supported by regular visits by other Florentines, and

Gatto began leading a group again, as revealed in this memory of Feruglio: "...The years spent in Geneva, where I could devote most of my time to research, collaborating with a very active group, were among the best of my professional life. I arrived in Geneva after my military service, during which my research activity had been interrupted for more than a year. Gatto was understanding when I first arrived, and gave me a relatively simple problem to solve, allowing me to recommence my research activity with some confidence. Gatto's ability to find new problems within the reach of young and inexperienced collaborators was exceptional. At the same time I was rather surprised at the way he interacted with the rest of the group. We, his collaborators, had our offices in a building which was detached from the School of Physics at the University of Geneva, where Gatto had his office. Despite the short distance, Gatto never came to our building. He preferred to talk to us over the phone on a daily basis. Every afternoon at around five o'clock, he called one of us to discuss the latest developments in the project we were working on. He would spend an hour or more bringing himself up to date but also making comments, asking questions and giving advice. The job of answering Gatto's penetrating and insistent questions was left to the oldest group member, Roberto Casalbuoni, who patiently and very effectively met his requests. Generally, once a week, the whole group migrated to Gatto's room for a discussion at the blackboard, in which even some of the younger members dared to present some points and some new ideas. These ideas were immediately abandoned if Gatto pronounced the famous words: "Very interesting, I've never seen anything like that in my entire career...". At that point we knew that we were completely off the mark. So, very kindly but very firmly, Gatto stimulated us and directed us towards the realisation of the project we were working on."

To this memory of Feruglio on how the group was organised we have a comment by Casalbuoni to add: "Gatto had an extraordinary ability to identify new research projects. He generally tried to consider new proposals that emerged in literature. A typical problem concerned the models composed for quarks and leptons, on which we started to work immediately after Harari's first proposal. Something similar happened with the Higgs boson studies, when the interest in this particle was stimulated by the announcement of DESY, in 1984, that a particle with a mass of 8.33 GeV had been discovered in Doris. This discovery was not confirmed by other experiments, nor by DESY, but it was important because both theorists and experimentalists were stimulated to deal with the Higgs issue, which had been somewhat neglected until then. Gatto's sixth sense was proverbial among us, not only at research level, but also for more banal things. When the conceptual aspects of a problem were solved and there were only calculations to be made, Gatto left us alone and made himself scarce. But incredibly, as soon as we had finished the calculations, the phone rang, it was Gatto...".

The effective model that the group had proposed for describing a heavy Higgs was based on the theory of non-linear symmetries and was equivalent to imposing a constraint on the Higgs' field. In 1988, Gatto had proposed the study of an analogous phenomenon in supergravity, imagining the possibility of a heavy gravitino. This study allowed the formulation of an equivalence theorem, similar to the one valid for ordinary gauge theories. In 1989, these ideas were applied by the group to a supersymmetric theory in which the scalar was heavy, finding that, in this situation, the superfield satisfies the constraint that its square must be zero. These studies have recently been applied to the non-linear realisations of supergravity.

At more or less the same time, Gatto's collaboration with Barducci, Casalbuoni, De Curtis and Giulio Pettini, another student of Casalbuoni, tackled the study of QCD at a finite temperature and density. These studies are relevant in many hadronic physics problems. The most important result, found in this research, was that in the phase diagram as a function of baryonic number and temperature, the existence of a tricritical point was possible, this being a point where a first order phase transition line ends in a second one.

Casalbuoni remembers this result: "What we had found was completely unexpected, no one in literature had mentioned such a possibility. At first we thought it was an effect due to the approximations used. We then decided to calculate the phase diagram of the Gross and Neveu model, a two-dimensional model which has many of the characteristics of QCD. We tested the existence of a tricritical point in this model too. The Aachen Conference was held in September 1989, with a session dedicated to these issues. I remember that my presentation was met by palpable scepticism. Karsch, an expert on lattice calculations, found it particularly hard to believe. It must be said, to his credit, that the calculations on the grid were made at zero density, due to technical problems, while the tricritical point is an effect of finite density. We later discovered that two Japanese, Asakawa and Yazaki, had achieved the same result using a different approximation of QCD. Now, despite the lack of experimental data, the widespread opinion is that the tricritical point in QCD really exists, even for partial results obtained on the lattice".

But at that time, between the end of 1989 and the beginning of 1990, the LEP at CERN and its competitor, SLC at SLAC, came into operation. The group immersed themselves in much more phenomenological work. Peskin and Takeushi before, Altarelli, Barbieri and Caravaglios after, had shown that the characteristics of the vector boson Z, which was to be studied with great precision at the two machines, could be summarised in three parameters, whose deviations from the zero value represented a measure of the effects of a possible new physics beyond the Standard Model. The group focused on the deviations of these parameters due to the effective model for a heavy Higgs, which had been introduced earlier, called BESS (Breaking Electro-weak Symmetry Strongly) by Feruglio. Part of these works and also the study of heavy vector bosons were carried out not only by the group made up of Gatto, Casalbuoni, De Curtis, Dominici and Feruglio, but also by Guido Altarelli and Pierre Chiappetta.

In 1992, two of Gatto's PhD students, Aldo Deandrea and Nicola di Bartolomeo arrived in Geneva, and began studying the problems related to the BESS model, before becoming involved, thanks to a collaboration begun the year before with Giuseppe Nardulli from Bari (deceased in 2008), in a line of research on effective models for heavy mesons. This line of research ended with a review, published in Physics Report in 1997, which has been quoted numerous times since.

In the late 90s, when the LEP was close to closure and about to be replaced by LHC, Gatto and collaborators resumed working on the heavy Higgs issue. Given the strict limits which LEP had set on possible deviations from the Standard Model, they developed a model (degenerate BESS) which would work both in linear and non-linear formulations.

At that time, two other Florentines, Michele Modugno, one of Casalbuoni's students, and Lapo Casetti, who had been a student of Marco Pettini, collaborated in research on finite temperature QCD and on statistical systems.

In 1975, it was shown that at high densities and zero temperature, colour symmetry could be broken with a mechanism similar to that of superconductivity, with the formation of a diquark condensate. The calculations of the time, however, resulted in a gap of a few MeV, and, therefore, of little phenomenological interest. In 1998, these calculations were called into question and it was shown that, contrary to what had previously been found, the gap could go up to hundreds of MeV. Since the group had worked a lot on QCD at finite temperature and density, Gatto immediately showed interest in these researches and when Casalbuoni visited CERN in July 1999, they began studying this issue. The first and immediate outcome was the formulation of an effective model, based on a scheme similar to that describing the breaking of chiral symmetry in QCD, which defined the Goldstone bosons associated with the breaking of colour symmetry. This model has become a classic for this kind of physics. In the summer of 2000, Casalbuoni and Nardulli were both at DESY, and so yet another collaboration was born with Gatto, Nardulli and his students from Bari. Many aspects of high density QCD, with various possible phases, were studied in this project. The most interesting result was the discovery of an instability in the phase which was deemed the most energetically favourable. In a subsequent work, it was shown that such instability would not occur in a non-homogeneous phase: the so-called LOFF (Larkin-Ovchinnikov-Fulde-Ferrel) phase, extensively studied by physicists of the structure of matter.

The almost thirty-year collaboration with Casalbuoni ended in 2006 due to the latter's involvement in the activities of the Galileo Galilei Institute (GGI) in Florence, which became operational that same year. The only exception was a review on non-homogeneous phases of QCD, which ended in 2014. Nardulli died prematurely in 2008 but Gatto continued collaborating with his former-students until 2014, when his conditions detriorated no longer allowing him to work.

It became increasingly difficult to communicate with Gatto after 2014. The fact that he could no longer continue his activity as researcher had a profound effect on him, considering how much of his time he had dedicated to it. Gatto tried every day to read the most interesting articles that would come out, in order to find new topics to suggest to his collaborators, choosing the most interesting and those that could be dealt with most easily by the group, keeping in mind the professional experiences of the various people. His interest in research was his whole life, and he carried it on until the end. Stefania De Curtis tells us: "The last time I met Raoul was at Cern in 2009 during the workshop "From the LHC to a Future Collider". It had been several years since I had last seen him, and I was truly surprised to see him arrive, quietly, with his well-known "catlike footstep", at the beginning of my seminar. He sat in one of the side seats and remained so quiet and attentive. At the end of the meeting I approached him to say hello and he did not fail to compliment me and show his usual curiosity and interest in what I was working on. I realised how, although we were no longer collaborating, he had monitored the evolution of our line of research in every detail. His comments were as detailed and enlightening as ever. I can't deny that I was amazed, despite having known Gatto for a long time. I remember that, when he said goodbye, he added "Blimey! How much progress you've made!" with a note of regret."

We would also like to point out that in 1975 Gatto gained an important acknowledgement for his activity as researcher and "maestro", receiving the Award of the President of the Italian Republic (see Fig. 2). In 2003, he also received the Enrico Fermi Prize, awarded by the Italian Physics Society for the following reason: "...for his pioneering work in the field of weak decays of strange particles and for his role as leader in a fundamental field of subnuclear physics.

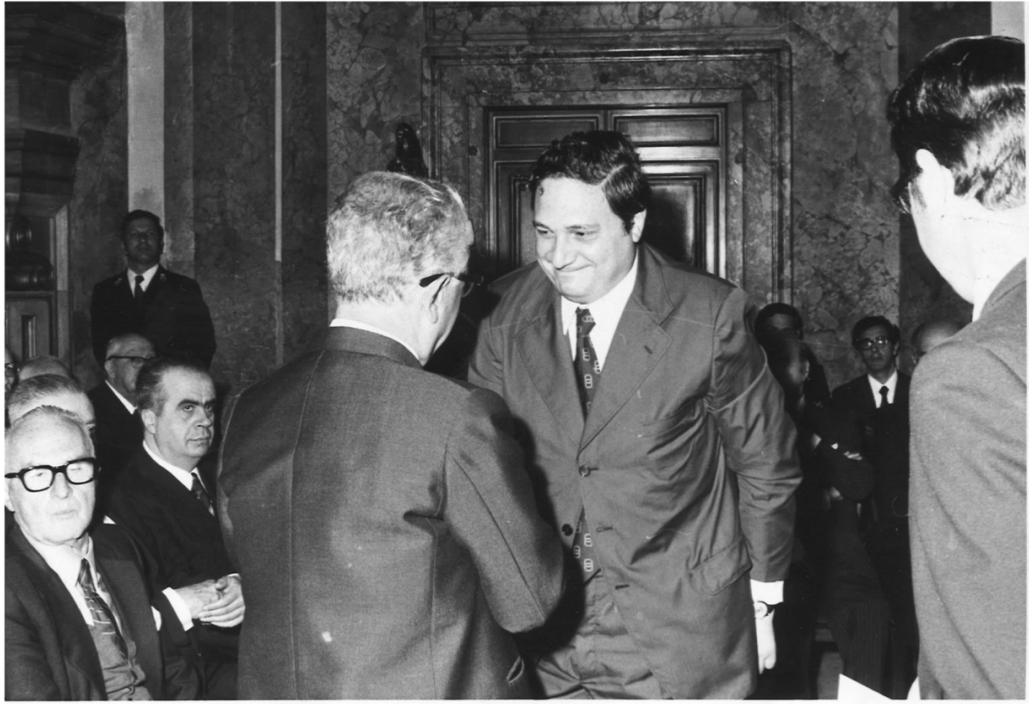

Fig. 2 - *Raoul Gatto receives the Award of the President of the Italian Republic*

Gatto's lessons were always extremely clear, but his teaching was particularly effective when he oriented younger collaborators towards a particular research job, which he almost always managed to choose in such a way that it could be completed. He was very capable of estimating both the difficulty of the job and the ability of his collaborators to carry out research.

In 1990, the Department of Theoretical Physics of the University of Geneva organised a conference to celebrate the 60$^{th}$ birthday of Gatto and Henri Ruegg. On the 17$^{th}$ of January 1997, again the Department of Theoretical Physics at the University of Geneva, celebrated Gatto's retirement with the "Theoretical Physics Conference in Honour of Raoul Gatto", in the presence of many of his former students and with speeches by Altarelli, Cabibbo, Ferrara, Gallavotti and Maiani. On this occasion, we also had the pleasure of meeting his wife "Reginetta", whom Gatto was often forced to phone during our meetings, which would last longer than expected, to tell her he'd be late.

Finally, here are some portraits of Gatto by his collegues: Carlo Bernardini [Bernardini, 2006] saw Gatto as "A strange loner, admired by his pupils". Without a doubt, Gatto had a very shy character. At social events at the Department of Theoretical Physics in Geneva for example, he always arrived one minute before the beginning and left as soon as the social event was over. But we spent entire afternoons, evenings and late nights together, having pleasant discussions and thinking about how to solve the problems raised by our scientific work." Here's the description presented by Preparata [Preparata, 2002] when, in 1963, he went to Frascati to meet Gatto. "Although he was quite young – he must have been just over thirty- Gatto was considered one of the world's top physicists. Gatto was a big, slightly overweight boy, with an open and friendly face, maybe a bit shy; I liked him immediately."

With Gatto's death, Italian theoretical physics lost a great teacher and a world-renown great scientist.

M. Greco, G. Pancheri, Analysis, 2008, p. 21
C. Bernardini, Fisica vissuta, Hoepli, 2006
G. Preparata, Dai quark ai cristalli. Breve storia di un lungo viaggio dentro la materia.
Bollati Boringhieri, 2002
Gatto's publications are available on the website:
http://inspirehep.net/search?ln=it&ln=it&p=a+gatto%2C+r&of=hb&action_search=Cerca&sf=earliestdate&so=d&rm=&rg=25&sc=0

Other recent papers on R. Gatto:
L. Maiani, Raoul Raffaele Gatto: an unforgettable Maestro , Il Nuovo Saggiatore 2018;
G. Battimelli, F. Buccella, P. Napolitano, Raoul Gatto, a great Italian scientist and teacher in theoretical elementary particle physics , Giornale di Fisica, 2018

Roberto Casalbuoni is Professor of Theoretical Physics at University of Florence. He has held numerous management positions at INFN and University of Florence. His research activity is in the field of Theoretical Physics of Elementary Particles. He worked with Raoul Gatto for almost thirty years.

Daniele Dominici is Professor of Theoretical Physics at University of Florence, and Director of the National Centre for Advanced Studies  Galileo Galilei Institute for Theoretical Physics. He carries out research activities in the field of Theoretical Physics of Elementary Particles. He worked with Raoul Gatto for many years, most notably a sabbatical at the University of Geneva in 1984-1986.